\documentclass[%
aps,
amsmath,amssymb,
reprint,
superscriptaddress
]{revtex4-2}

\usepackage{graphicx}
\usepackage{dcolumn}
\usepackage{hyperref}
\hypersetup{
	breaklinks=true,
	colorlinks=true,
	linkcolor=[RGB]{40,42,143},
	filecolor=[RGB]{40,42,143},
	citecolor=[RGB]{40,42,143},      
	urlcolor =[RGB]{40,42,143},
}

\newcommand{\appsection}[1]{\section{\MakeUppercase{#1}}}
\newcommand{\bx}  {\boldsymbol{x}}
\begin{document}
	
	
	\title{Impact of wavefront shape on nonlinear ultrasound imaging of monodisperse microbubbles}
	
	\author{Agisilaos Matalliotakis}
	\affiliation{Department of Imaging Physics, Faculty of Applied Sciences, Delft University of Technology, 2628 CJ Delft, the Netherlands}
	
	\author{Rick Waasdorp}
	\affiliation{Department of Imaging Physics, Faculty of Applied Sciences, Delft University of Technology, 2628 CJ Delft, the Netherlands}
	
	\author{Martin D. Verweij}%
	\affiliation{Department of Imaging Physics, Faculty of Applied Sciences, Delft University of Technology, 2628 CJ Delft, the Netherlands}
	\affiliation{Department of Biomedical Engineering, Erasmus University Medical Center, 3000 CA Rotterdam, the Netherlands}
	
	\author{David Maresca}%
	\email{Corresponding author: d.maresca@tudelft.nl}
	\affiliation{Department of Imaging Physics, Faculty of Applied Sciences, Delft University of Technology, 2628 CJ Delft, the Netherlands}
	
	
	\begin{abstract}
		The field of contrast-enhanced ultrasound (CEUS) combines nonlinear, resonant microbubbles (MBs) with dedicated pulse sequences to reveal the vascular function of organs. Clinical ultrasound contrast agents consist of polydisperse MB suspensions with diameters ranging from 0.5 to 10 µm and resonance frequencies ranging from 1 to 15 MHz. As a result, just a small fraction of MBs resonates at a given ultrasound frequency. MB suspensions with narrow size distributions can be tuned for a specific imaging frequency, boost CEUS sensitivity and enable deeper vascular imaging. However, their enhanced nonlinear behavior makes imaging susceptible to nonlinear wave propagation artifacts. Here, we numerically investigate the impact of the acoustic wavefront shape on the  imaging of nonlinear, monodisperse MBs. Specifically, our approach relies on an extension of the Iterative Nonlinear Contrast Source (INCS) method that accounts for all nonlinear effects in CEUS. We demonstrate that supersonic x-shaped wavefronts referred to as x-waves can be used to generate ultrasound images of monodisperse MBs without nonlinear wave propagation artifacts. On the contrary, imaging based on focused, planar and diverging wavefronts leads to significant nonlinear artifacts. Taken together, our results show that x-waves can harness the full potential of monodisperse MBs by enabling their sensitive and specific detection in a tissue context.
		
	\end{abstract}
	
	\maketitle
	
	
	\section{Introduction} \label{sec:Introduction}
	
	The combination of ultrasound imaging with intravenously administered echogenic MBs has enabled significant advances such as microvascular imaging beyond the diffraction limit \cite{Errico2015}. MBs are nonlinear oscillators that scatter ultrasound efficiently thanks to their high contrast in density and compressibility with blood plasma \cite{Heiles2021} and their resonant behavior in the MHz range \cite{Marmottant2005}. Ultrasound pulse sequences \cite{Averkiou2020} exploit nonlinear MB scattering to detect these vascular agents deep into tissues. A drawback of established pulse sequences such as amplitude modulation (AM) imaging is their susceptibility to cumulative nonlinear effects induced by wave propagation in MB suspensions \cite{Tang2006,tenKate2012}. Briefly, ultrasound waves propagating in a highly nonlinear effective medium such as a resonant MB suspension experience amplitude-dependent attenuation and amplitude-dependent speed of sound variations \cite{Emmer2009, Sojahrood2023}. As a result, waves with different amplitudes get distorted differently through a MB suspension. In the end, waves carry that nonlinear distortion deeper into the medium, beyond the MB suspension, where they cause echoes that make AM imaging to misclassify tissue as MBs.
	
	In 2015, Renaud et al. \cite{Renaud2015} demonstrated that AM imaging based on the intersection of two diverging wavefronts could improve CEUS specificity. Maresca, Sawyer et al. \cite{Maresca2018} further optimized this approach into a sequence called cross-amplitude modulation (xAM) that relies on cross-propagating plane waves intersecting with a constant angle. xAM presents fundamental advantages over previous AM implementations. First, the constant cross-propagation angle ensures a constant reduction of cumulative nonlinear effects along the line where the wavefronts cross. Second, it generates a non-diffractive beam pattern with a constant beam width \cite{Nagai1982,Szabo2014} that improves the lateral resolution of ultrasound images.
	
	This study numerically investigates the specificity of AM imaging of monodisperse MBs, the most nonlinear ultrasound contrast agent to date \cite{Segers2018}, for implementations using focused \cite{Averkiou2020}, planar \cite{Tremblay2015}, diverging \cite{Yan2023} and x-shaped \cite{Maresca2018} ultrasonic wavefronts. We use the Iterative Nonlinear Contrast Source (INCS) method \cite{Koos_Thesis,INCS} to solve the full nonlinear acoustic wave equation in an acoustic medium containing a suspension of resonant monodisperse MBs. INCS was extended to account for ultrasound attenuation, medium inhomogeneities \cite{Libe_Thesis,INCS_Inhom}, local medium nonlinearities \cite{LocalNL2023} and nonlinearities arising from the oscillatory behavior of resonant MBs surrounded by tissue-mimicking linear scatterers. \cite{Bubble_Cloud} Together, these capabilities allow us to simulate CEUS imaging modes such as AM and xAM.
	
	The manuscript is organized as follows. Secs.~\ref{sec:Incident} and~\ref{sec:ContrastSources} describe the INCS method and its application to AM imaging. Sec.~\ref{sec:ContrastDomain} describes the media used in the simulations. Sec.~\ref{sec:Beamforming} describes the beamforming process that has been implemented to reconstruct ultrasound images. Sec.~\ref{subsec:Results_IncidentField} presents the incident and residual acoustic pressure fields generated by AM pulses in a homogeneous nonlinear medium free of ultrasound contrast agents. Sec.~\ref{subsec:Results_TotalFields} presents the total and residual acoustic pressure fields in a medium containing a monodisperse MB suspension. Sec.~\ref{subsec:Results_Supersonic} describes the origin of nonlinear wave propagation artifacts observed in AM imaging. Furthermore, Sec.~\ref{subsec:Results_Imaging} presents numerical AM images of monodisperse MBs generated with x-shaped, focused, planar and diverging wavefronts. Finally, a brief analysis of the generated results is presented in Sec.~\ref{sec:Discussion}, while concluding remarks of this study are given in Sec.~\ref{sec:Conclusions}.
	%
	\section{Methods} \label{sec:ProblemForm}
	%
	\subsection{Extension of INCS to account for nonlinear effects in CEUS}\label{sec:ContrastSources}
	The linear incident pressure field resulting from an external source in a linear, lossless, homogeneous and isotropic background medium can be mathematically described by the wave equation
	\begin{equation}
		c^{-2}_0\partial^{2}_t p-\nabla^2 p= S_\mathrm{pr},
		\label{eq:LosslessLinWestervelt}
	\end{equation}
	where $p$ [Pa] represents the acoustic pressure. The parameter $c_{0}$ [m/s] denotes the sound speed in the background medium. The operator ${\nabla}^2$ denotes the Laplacian and $\partial^{2}_t$ represents the second order time derivative. On the right-hand side of the equation, the primary source term $S_\mathrm{pr}$ is used to describe the action of the transducer. The linear acoustic pressure distribution arising from the primary source in the background medium can be explicitly found as
	\begin{equation}
		p(\bx,t) = G(\bx,t)\ast_{\bx,t}S_\mathrm{pr}(\bx,t),
		\label{eq:background_solution}
	\end{equation}
	where $G(\bx,t)$ is the Green's function of the background medium, and the operator $\ast_{\bx,t}$ denotes convolution over the spatiotemporal domain of $S_\mathrm{pr}$. Because the background medium is simple, $G(\bx,t)$ is known analytically.
	
	Realistic media are modeled by more complex wave equations that also account for inhomogeneities, attenuation, nonlinear behavior, etc. Within the doctrine of the INCS method, all terms in which the more complex wave equation deviates from the simple background wave equation in Eq.~(\ref{eq:LosslessLinWestervelt}) are shifted to the right hand side and are considered as contrast sources. The result is
	\begin{equation}
		c^{-2}_0\partial^{2}_t p-\nabla^2 p= S_\mathrm{pr}+S_\mathrm{cs}(p),
		\label{eq:waveeq_cs}
	\end{equation}
	where $S_\mathrm{cs}(p)$ is the total of all contrast source terms that account for the inhomogeneity, attenuation, nonlinearity, and the like that are not exhibited by the background medium. The acoustic pressure distribution arising from the primary source in a realistic medium can be implicitly found as
	\begin{equation}
		p(\bx,t) = G(\bx,t)\ast_{\bx,t}[S_\mathrm{pr}(\bx,t)+S_\mathrm{cs}(p)].
		\label{eq:fullwave_solution}
	\end{equation}
	With the INCS method, an explicit solution is obtained in an iterative way, e.g. by using the Neumann iterative scheme
	\begin{align}
		p^{(0)}&=G(\bx,t)\ast_{\bx,t}S_\mathrm{pr}(\bx,t),\label{eq:neumann_scheme_p0}\\
		p^{(n)}&=G(\bx,t)\ast_{\bx,t}[S_\mathrm{pr}(\bx,t)+S_\mathrm{cs}(p^{(n-1)})]\nonumber\\
		&=p^{(0)}+G(\bx,t)\ast_{\bx,t}S_\mathrm{cs}(p^{(n-1)})],\;\;(n\geq 1).
		\label{eq:neumann_scheme}
	\end{align}
	This general approach is used for the numerical simulations in this paper.
	
	To simulate CEUS imaging, all nonlinear effects occurring during ultrasound wave propagation in a nonlinear medium containing nonlinearly scattering MBs must be taken into account. In addition, simulations should support scattering from tissue-mimicking linear scatterers (LSs) that typically surround a MB suspension in biological tissues. Based on these criteria, INCS was extended using extra contrast source terms resulting in the new nonlinear wave equation 
	\begin{align}
		\label{eq:LosslessContrast}
		{c^{-2}_{0}}\frac{\partial^{2} p}{\partial t^2}-\nabla^{2}p &= S_\mathrm{pr} + S_\mathrm{MB}(p) + S_\mathrm{LS}(p)\\ \nonumber
		& + S_\mathrm{globalNL}(p) + S_\mathrm{localNL}(p),
	\end{align}
	where $S_\mathrm{MBs}$ is the contrast source term describing the scattering of the MBs in the suspension \cite{Bubble_Cloud}, $S_\mathrm{LS}$ is the contrast source term describing the scattering of the linear scatterers in the LS suspension \cite{Bubble_Cloud}, $S_\mathrm{globalNL}$ and $S_\mathrm{localNL}$ are the terms describing global \cite{Bubble_Cloud} and local \cite{LocalNL2023} medium nonlinearities, respectively. The sum of all the contrast source terms yields the term $S_\mathrm{cs}(p)$ in Eq.~(\ref{eq:waveeq_cs}), which is then convolved with the 4D spatiotemporal Green's function of the linear background medium to compute a nonlinear field correction. Based on the Neumann iterative scheme described in Eqs.~(\ref{eq:neumann_scheme_p0}) and (\ref{eq:neumann_scheme}), INCS successively generates an increasingly accurate solution of Eq.~(\ref{eq:LosslessContrast}), taking into account nonlinear effects and multiple scattering.
	%
	\subsection{INCS implementation of AM pulse sequences} \label{sec:Incident}
	\begin{figure*}[htbp!]
		\centering
		\includegraphics{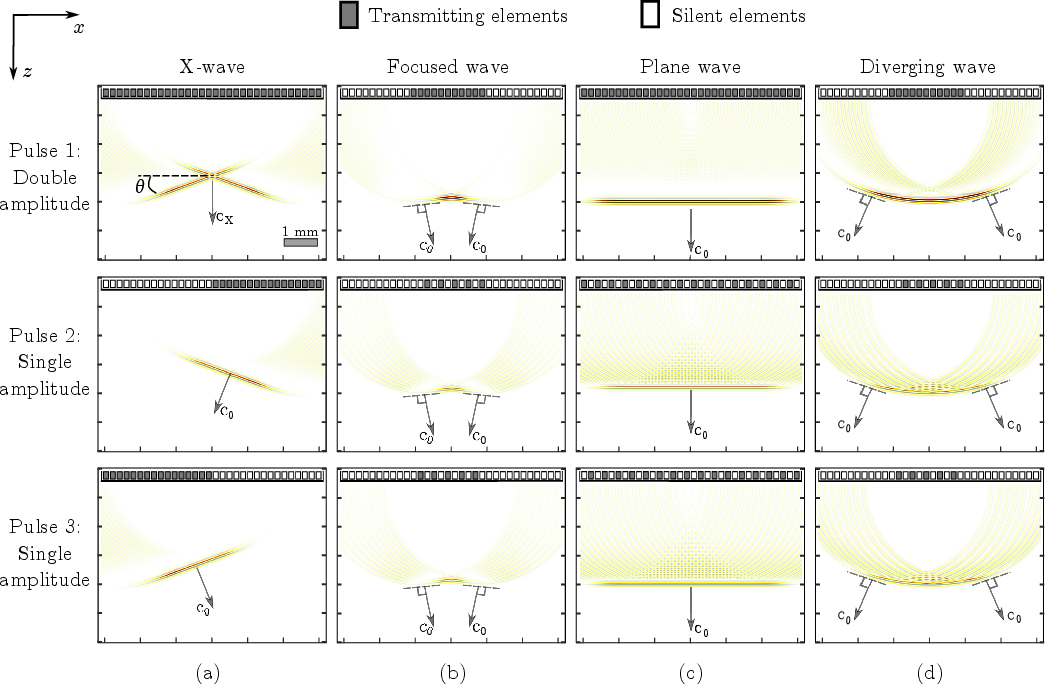}
		\caption{Wavefront shapes belonging to the pulse sequences investigated for AM ultrasound imaging of monodisperse MBs. (a) xAM sequence with a wavefront angle $\theta=20^{\circ}$, (b) focused AM sequence, (c) plane wave AM sequence and (d) diverging AM sequence. The first row shows the imaging pulses transmitted to elicit a double amplitude response from the medium, while the second and third rows show the pulses transmitted to elicit a single amplitude from the medium. Arrows indicate wavefront velocities.}
		\label{fig:ImagingSequences}
	\end{figure*}
	In this study, we simulate a 64-element ($N_\mathrm{tr}=64$), 15 MHz linear transducer array emulating a L22-14vX probe (Verasonics, Kirkland, WA, USA). Transducer elements have a height $H_{\mathrm{el}} = 1.4\ \mathrm{mm}$, a width $W_{\mathrm{el}}=0.08\ \mathrm{mm}$, a pitch $P_{\mathrm{tr}}=0.1\ \mathrm{mm}$ and the total focal distance of the array, which includes a lens in elevation is $z_\mathrm{el}=8\ \mathrm{mm}$. The natural focus of the transducer array is $(N_\mathrm{tr}\,P_{\mathrm{tr}})^2/4\lambda=103$ mm in the azimuthal plane and  $H_{\mathrm{el}}^2/4\lambda=5\ \mathrm{mm}$ in the elevational plane, with $\lambda$ being the wavelength. All transmitted ultrasound waveforms used in this study are Gaussian-windowed sine bursts defined as follows
	\begin{equation}
		s(t)=\mathrm{exp}\left[-\left(\frac{t-T_\mathrm{d}}{T_\mathrm{w}/2}\right)^2\right]\mathrm{sin}[2\pi f_0 (t-T_\mathrm{d})],
		\label{eq:Time_Signature}
	\end{equation}
	where $f_0 = 15\;\mathrm{MHz}$ is the center frequency, $T_w=1.5/f_0=0.1\ \mu$s is the width of the Gaussian envelope and $T_d=3/f_0=0.2\ \mu$s is the temporal delay of the window. The simulated medium is water, characterized by a density of mass $\rho_0 = 1060\;\mathrm{kg/m^3}$ and a speed of sound $c_0 = 1482\;\mathrm{m/s}$.
	
	The first AM sequence implemented in INCS is the sequence with x-shaped wavefronts. \cite{Maresca2018}. xAM pulse transmissions proceed as follows: first, elements 1 to $N/2$ and elements $N/2+1$ to $N$ of the array simultaneously transmit two axisymmetric, tilted plane waves with wavefronts that make an angle $\theta$ and -$\theta$ with respect to the array (see pulse 1, Fig.~\ref{fig:ImagingSequences}(a)). Second, only elements 1 to $N/2$ transmit a single tilted plane wave at an angle -$\theta$ with respect to the array  (see pulse 2, Fig.~\ref{fig:ImagingSequences}(a)). Third, only elements $N/2+1$ to $N$ transmit a single symmetric plane wave at an angle $\theta$ with respect to the array (see pulse 3, Fig.~\ref{fig:ImagingSequences}(a)). The two cross-propagating waves depicted as pulse 1 interfere in a small zone around a virtual bisector that separates the two half-apertures. Particles of the insonified medium that are positioned along the bisector experience an identical pressure amplitude from pulses 2 and 3, but a double pressure amplitude from pulse 1 (x-wave transmission). Note that the overpressure at the plane waves' intersection depicted in pulse 1 propagates along the bisector with a supersonic phase velocity
	\begin{equation}
		c_\mathrm{X} = c_0/\cos(\theta).
	\end{equation}

	The second AM sequence implemented in INCS relies on focused ultrasound wavefronts as depicted in Fig.~\ref{fig:ImagingSequences}(b). Focused AM transmissions proceed as follows: first, all elements of the array are activated following a parabolic delay law to generate a high amplitude transmission (see pulse 1, Fig.~\ref{fig:ImagingSequences}(b)). Second, even transducer elements of the same aperture are activated using the same delay law to generate a half amplitude transmission (see pulse 2, Fig.~\ref{fig:ImagingSequences}(b)). Third, odd transducer elements of the same aperture are activated using the same delay law to generate a second half amplitude transmission (see pulse 3, Fig.~\ref{fig:ImagingSequences}(b)). Here the focal distance is set to $z_\mathrm{f}=5$ mm. Note that in focused AM, acoustic wavefronts propagate at speed of sound $c_0$.
	
	The third AM sequence implemented in INCS relies on planar ultrasound wavefronts as seen in Fig.~\ref{fig:ImagingSequences}(c). Plane wave transmissions are generated by transmitting the same pulse with all elements of the transducer array. Here as well, the transmitted acoustic pressure amplitude is modulated using either all, even or odd elements of the array aperture (Fig.~\ref{fig:ImagingSequences}(c)). Transmitted plane waves propagate at speed of sound $c_0$ in the $z$ direction.
	
	The fourth AM sequence implemented in INCS relies on diverging acoustic wavefront transmissions. A virtual point source located at $(x_\mathrm{f}, z_\mathrm{f})=(0, -3)$ mm is used to generate the diverging delay law. Here as well, the transmitted acoustic pressure amplitude is modulated using either all, even or odd elements of the array aperture (Fig.~\ref{fig:ImagingSequences}(d)). Transmitted diverging waves also propagate at speed of sound $c_0$.	
	\subsection{Simulation of a monodisperse MB suspension} \label{sec:ContrastDomain} 
	\begin{figure}[tbp!]
		\includegraphics{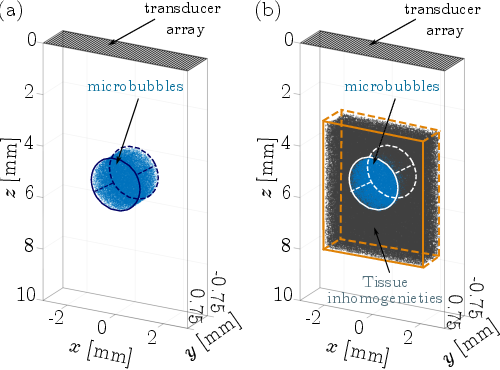}
		\caption{Media simulated in INCS. (a) Computational domain containing a monodisperse MB concentration equal to 5.7 $\times 10^6$  ml$^{-1}$ (blue) of 0.6 $\mu$m radius bubbles, embedded in water. (b) Configuration incorporating a tissue-mimicking scatterer concentration of 6.2 $\times 10^6$ ml$^{-1}$ (grey) surrounding the monodisperse MB suspension in (a). }
		\label{fig:ContrastDomain}
	\end{figure}
	The dimensions of the computational domain are set to $X \times Y \times Z = 6.4\ \mathrm{mm} \times 1.5\ \mathrm{mm} \times 10\ \mathrm{mm}$ as illustrated in Fig.~\ref{fig:ContrastDomain}. To accurately solve the full nonlinear wave equation up to the second harmonic frequency ($h=2$), we choose a Nyquist frequency $F_{nyq}$ equal to at least $h+1.5=3.5$ times of the center ultrasound frequency. This corresponds to a sampling frequency of $F_s=2\cdot F_{nyq}=2\cdot3.5\cdot 15=105$ MHz as a basis for discretizing the spatiotemporal domain. Furthermore, it is know that we need at least $j=h+1=3$ iterations for an accurate prediction of the second harmonic \cite{INCS}. To achieve a relative root mean square error of $10^{-6}$ between successive iterations below, we take $j=5$ iterations. Consequently, our simulations account for MB interactions up to 4th order multiple scattering \cite{Bubble_Cloud}.
	
	We consider a monodisperse MB suspension confined in a cylindrical volume positioned at the center of the simulation domain $(x,\ y,\ z) = (0,\ 0,\ 5)$ mm, with a radius of 1 mm in the $xz$ plane and a length of 1.4 mm in the $y$ dimension, as shown in Fig.~\ref{fig:ContrastDomain}(a). This geometry is chosen to mimic MBs filling the lumen of a blood vessel. A MB radius $R=0.6\ \mu\mathrm{m}$, corresponding to a resonance frequency of 15 MHz in the elastic oscillation regime, is selected to match the center frequency of the simulated transducer array. We use the Marmottant model \cite{Marmottant2005} to emulate the nonlinear behavior of each MB in the simulation domain. Simulated material properties of MBs are provided in Table~\ref{table:Marmottant} \cite{Segers2018a}. 
	\begin{table}[t!]
		\caption{\label{table:Marmottant}Material properties of a single simulated MB with $R=0.6\  \mu\mathrm{m}$. $\kappa_{s}$ [kg/s] is the shell viscosity, $\sigma_\mathrm{w}$ [N/m] is the surface tension of water, $\sigma_\mathrm{R}$ [N/m] is the effective surface tension, $\gamma$ is the polytropic exponent of the gas encapsulated in the bubble, $\chi$ [N/m] is the shell elasticity, $\mu$ [Pa$\cdot$s] is the gas core viscosity.}
		\centering                                
			\begin{tabular}{wc{1.4cm}wc{1.6cm}wc{1.6cm}wc{0.6cm}wc{1.4cm}wc{1.1cm}}
				\hline
				$\kappa_\mathrm{s}\;[\mathrm{kg/s}]$ & $\sigma_\mathrm{w}\;[\mathrm{N/m}]$ &$\sigma_\mathrm{R}\; [\mathrm{N/m}]$ & $\gamma$ & $\chi\;[\mathrm{N/m}]$ & $\mu\; [\mathrm{Pa}\cdot\mathrm{s}]$\\
				$2.4\times10^{-9}$ &  $0.072$ & $0.01$ & $1.07$ & 0.5  & $2\times10^{-3}$\\
				\hline
			\end{tabular}
		\end{table}

		A total of 25,000 monodisperse MBs are randomly distributed within the cylindrical volume, resulting in a MB concentration of $5.7\times10^6$ $\mathrm{ml}^{-1}$ and a gas volume concentration of $5.15\times 10^{-6}$, which follows injection protocols reported by MB manufacturers (e.g. Fujifilm VisualSonics, Toronto, ON, Canada).
		
		To reveal AM imaging artifacts due to nonlinear wave propagation, we also consider a second domain with tissue mimicking linear scatterers (Fig.~\ref{fig:ContrastDomain}(b)). These are located in a volume of $L\times W\times H= 4\ \mathrm{mm} \times 1.5\ \mathrm{mm} \times 5\ \mathrm{mm}$ surrounding the MB suspension. The linear scatterers have a spherical radius of 0.6 $\mu\mathrm{m}$ to match the speckle pattern generated by monodisperse MBs. Their concentration is $6.2\times10^6$ $\mathrm{ml}^{-1}$, and their volume concentration is $5.61\times 10^{-6}$, which is approximately similar to the MB suspension. The scattering coefficient of the linear scatterers \cite{Bubble_Cloud} is set to $g=6.57\times 10^{-6}$ m to match the scattering level of monodisperse MBs.
		\subsection{Ultrasound image reconstruction} \label{sec:Beamforming}
\begin{figure}[t!]
			\includegraphics{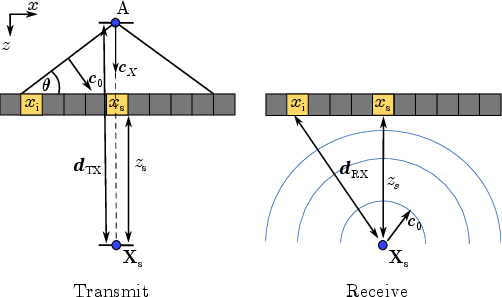}
			\caption{Transmit and receive geometry model used for xAM image reconstruction. }
			\label{fig:Sketch_BeamForming}
		\end{figure}

		%
		xAM and focused AM images are reconstructed line-by-line using 42 sliding sub-apertures across the MB suspension. For each sub-aperture position, 3 pulses are transmitted sequentially as illustrated in  Figs.~\ref{fig:ImagingSequences}(a) and (b), leading to a total of 126 ultrasound transmissions per AM image. To acquire RF data corresponding to each image line, instead of moving the transducer, we translate the position of the MB suspension laterally with steps $dx = 0.1$ mm, which is the pitch of the imaging array. The reconstructed image width ranges from $x=-2.1$ mm to $x=2.1$ mm.
		Plane wave and diverging wave AM images are reconstructed out of 3 pulse transmissions only, as each insonification captures a wide 2D field of view (Fig.~\ref{fig:ImagingSequences}(c) and (d)). B-mode images are reconstructed using the first pulse of each AM sequence.
		
		xAM image reconstruction is performed along the bisector as illustrated in Fig.~\ref{fig:Sketch_BeamForming}. The transmit distance $d_\mathrm{TX}$ and the transmit time of flight $\tau_\mathrm{TX}$ to a scatterer $\boldsymbol{X}_\mathrm{s}=(x_{\mathrm{s}}$,$z_{\mathrm{s}}$) are given by
		\begin{equation}
			d_\mathrm{TX}(\theta, z_\mathrm{s} )=D_\mathrm{h}\tan\left(\theta\right) + z_\mathrm{s},
			\label{eq:d_tx}
		\end{equation}
		\begin{equation}
			\tau_\mathrm{TX} = \frac{d_\mathrm{TX}}{c_\mathrm{X}},
			\label{eq:xwave_tau_tx}
		\end{equation}
		with $D_h = N_\mathrm{tr} P_\mathrm{tr} / 2$ being the length of the half aperture. The receive distance $d_\mathrm{RX}$ and receive time of flight $\tau_\mathrm{RX}$ are given by, 
		\begin{equation}
			d_\mathrm{RX}(\boldsymbol{X}_\mathrm{s})=\sqrt{ \left( x_\mathrm{s} - x_\mathrm{i} \right)^2 + z_\mathrm{s}^2}, 
			\label{eq:d_rx}
		\end{equation}
		\begin{equation}
			\tau_\mathrm{RX} = \frac{d_\mathrm{RX}}{c_0}, 
			\label{eq:xwave_tau_rx}
		\end{equation}
		with $x_i$ the i-th element position. Knowledge of transmit and receive time of flights enables the reconstruction of one image line using a conventional delay-and-sum (DAS) beamforming algorithm \cite{DAS2021}. By repeating the process for each transmitting sub-aperture position, a xAM image of the MB suspension is generated.
		
		Focused AM image reconstruction is performed line-by-line as well using usual transmit and receive time of flights. For plane wave and diverging wave image reconstruction, we use the MUST \cite{MUSTToolbox} toolbox and a virtual point source formulation as described by Perrot et al. \cite{DAS2021}. 
		\begin{figure*}[t!]
			\centering
			\includegraphics{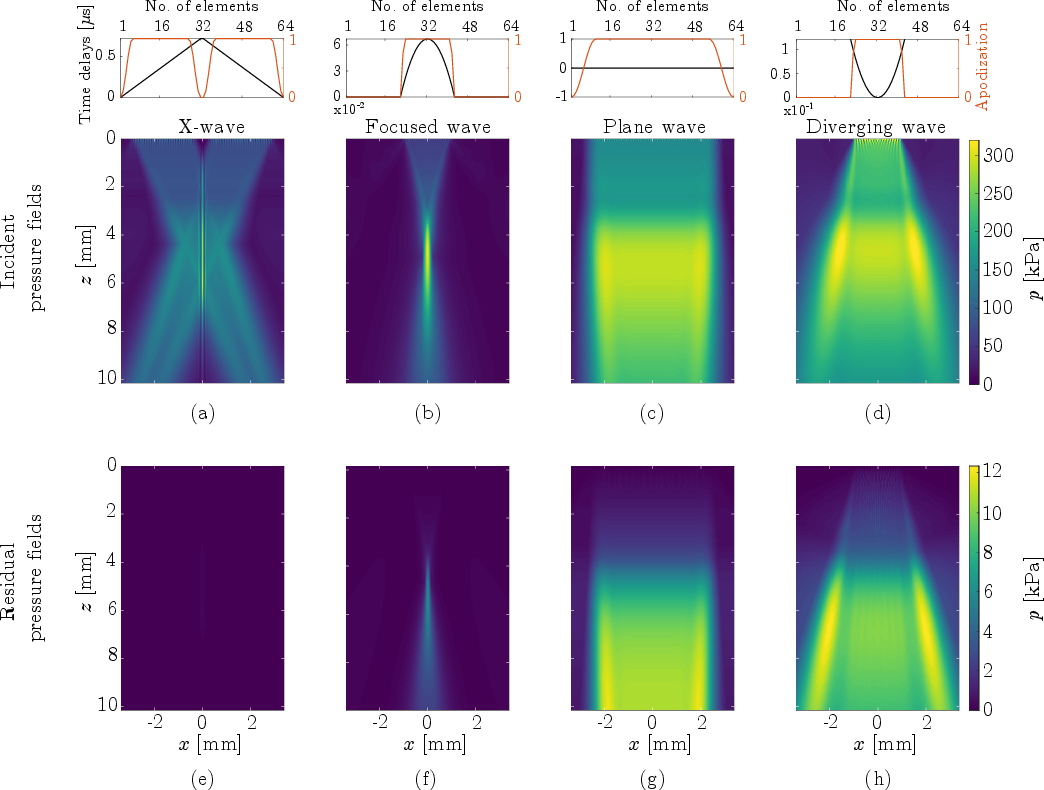}
			\caption{Acoustic pressure fields in a homogeneous nonlinear medium consisting of water. (Top row) Time delays (black) and apodization functions (orange) used for the first pulse transmissions. (a)-(d) Incident pressure fields generated by the first pulse transmission of each sequence using x-shaped, focused, planar and diverging beams, respectively.  (e)-(h) Residual AM pressure fields for the xAM, focused AM, planar AM and diverging AM pulse sequences, respectively. xAM transmissions are simulated for $\theta=20^{\circ}$.}
			\label{fig:IncidentFields}
		\end{figure*}
		%
		\section{Results} \label{sec:Results}
		%
		%
		\begin{figure*}[t!]
			\centering
			\includegraphics{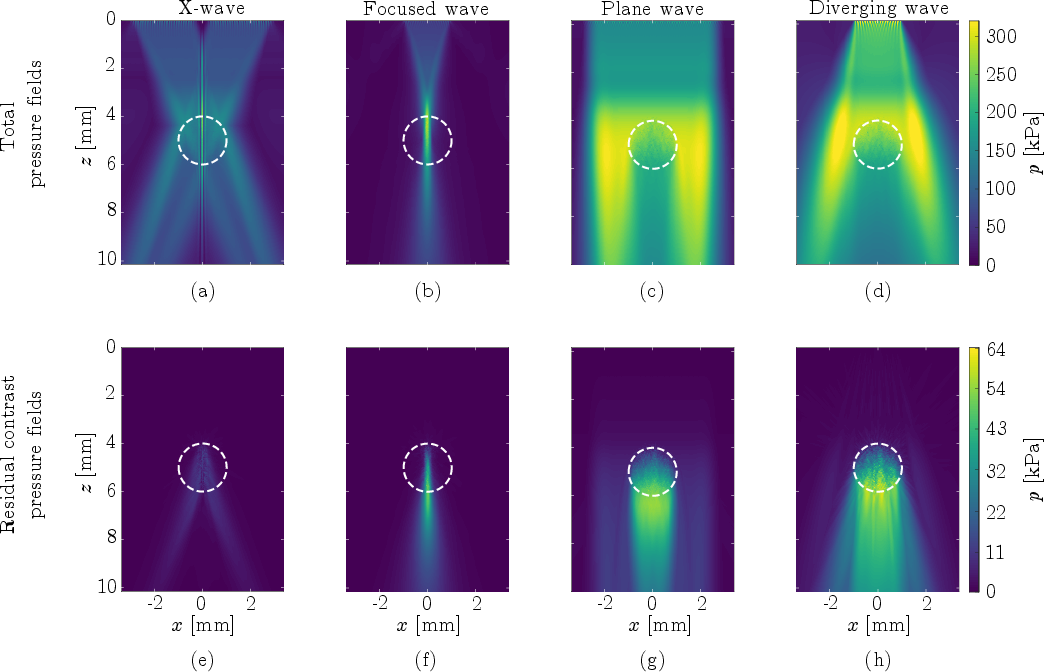}
			\caption{\label{fig:TotalPressureFields}
				Acoustic pressure fields in the presence of resonant monodisperse MBs. (a)-(d) Total acoustic pressure fields generated by the first pulse transmission of each sequence using x-shaped, focused, planar and diverging beams, respectively.  (e)-(h) Residual acoustic pressure fields generated by the xAM, focused AM, planar AM and diverging AM pulse sequences, respectively.}
		\end{figure*}
		%
		\subsection{Acoustic pressure fields in the absence of MBs} \label{subsec:Results_IncidentField}
		%
		To isolate nonlinear effects due to wave propagation, we first solve the full nonlinear wave equation in a homogeneous nonlinear medium consisting of water that is free of MBs.
		
		Fig.~\ref{fig:IncidentFields} presents the simulated incident pressure fields generated by the first pulse of each AM sequence, as well as the time delays and apodization functions used to generate these beam patterns. To ensure a fair comparison of nonlinear effects, we equated the average acoustic pressure delivered by the first pulse of each sequence on the $z$ axis between $z = 4\ \mathrm{mm}$ and $z = 6\ \mathrm{mm}$ (see Appendix ~\ref{app:PressureEq}). 
		
		We report in Fig.~\ref{fig:IncidentFields}(a) the x-wave beam pattern simulated for an angle $\theta = 20^{\circ}$. All 64 elements are used in transmission and a Tukey window apodization, with a parameter $\alpha=0.3$ representing the ratio of cosine-tapered section length to the entire window length, is applied along the elements of each half aperture of the array to mitigate the side lobe level and diffraction effects. The non-diffractive x-wave beam pattern exhibits a constant beam width along the bisector. Beyond the x-wave intersection zone, a noticeable acoustic pressure drop is observed along the bisector.
		
		Fig.~\ref{fig:IncidentFields}(b) displays the focused wave beam pattern generated with parabolic time delays applied to a sub-aperture of 22 elements apodized with a Tukey window ($\alpha=0.3$). The beam focus is equal to $z = 4.75\ \mathrm{mm}$, corresponding to an f-number of 2.27. Most of the focused beam energy is concentrated between $z = 4\ \mathrm{mm}$ and $z = 6\ \mathrm{mm}$. The focal length is shorter than the x-wave beam, whereas in the lateral direction, the focused beam is noticeably wider that the x-wave beam.
		
		Fig.~\ref{fig:IncidentFields}(c) displays the plane wave beam pattern generated with all 64 elements apodized with a Tukey window ($\alpha=0.3$). The plane wave beam focus is  $z = 5.25\ \mathrm{mm}$ and the elevated amplitude around $z=5\ \mathrm{mm}$ is due to the focusing in elevation.
		
		Fig.~\ref{fig:IncidentFields}(d) displays the diverging beam pattern generated by inverting the time delays used in Fig.~\ref{fig:IncidentFields}(b). Here the total beam focus is equal to $z = 4.65\ \mathrm{mm}$, also due to the focusing in elevation.
		
		The primary objective of CEUS imaging is to suppress nonlinear effects arising from wave propagation in the medium of interest, while retaining nonlinear effects arising from ultrasound contrast agents administered in that medium. To investigate the ability of AM sequences based on x-shaped, focused, planar and diverging wavefronts to suppress nonlinear wave propagation effects, we report the residual acoustic pressure fields after the AM operation (pulse 1 minus pulse 2 minus pulse 3) and display them on scale in Figs.~\ref{fig:IncidentFields}(e)-(h). 
		
		We observe that AM pulse sequences based on focused, planar and diverging wavefronts suffer from effects arising from nonlinear wave propagation and local noncollinear nonlinear interactions \cite{LocalNL2023}, even in the absence of MBs. The xAM residual pressure field on the contrary is nearly free of nonlinear effects. Quantitatively, the residual xAM pressure exhibits the lowest amplitude with a peak of 0.26 kPa (see Appendix ~\ref{app:ResidualFieldsInserts}, Fig.~\ref{fig:NonlinearInserts}(a)), followed by the residual focused AM pressure field with a peak of 6 kPa, the plane wave residual AM pressure field with a peak of 11.9 kPa, and the residual diverging AM pressure field with a peak of 12.3 kPa.
		
		Qualitatively, for the xAM case (Fig.~\ref{fig:NonlinearInserts}(a)), the highest residual pressure values are observed along the bisector. However, we observe that below the plane wave intersection zone the residual xAM pressure level drops significantly.
		
		For the focused AM case (Fig.~\ref{fig:IncidentFields}(f)), nonlinearities peak at the focus but extend beyond the focus because of cumulative nonlinear wave propagation effects, following the shape of the transmitted beam (Fig.~\ref{fig:IncidentFields}(b)).
		
		For the plane wave AM case (Fig.~\ref{fig:IncidentFields}(g)) a gradual onset of nonlinearities is observed as expected for cumulative nonlinear effects due to wave propagation \cite{Lai2019}. Additionally, a higher amplitude is observed on the sides of the plane wave beam due to the larger incident pressure in those regions.
		
		For the diverging AM case (Fig.~\ref{fig:IncidentFields}(h)), the behavior is similar to the plane wave case and nonlinearities reach the highest peak amplitude compared to all the other cases.
		\subsection{Acoustic pressure fields in the presence of monodisperse MBs} \label{subsec:Results_TotalFields}	
		%
		Total acoustic pressure fields generated by the first pulse of each AM sequence in the presence of a resonant monodisperse MB suspension are displayed in Figs.~\ref{fig:TotalPressureFields}(a)-(d). The x-wave pressure field (Fig.~\ref{fig:TotalPressureFields}(a)) is attenuated below the MB suspension and shows a peak acoustic pressure at $(x,\ z)=(0,\ \mathrm{mm},\ 7\ \mathrm{mm})$ of 166 kPa compared to 216 kPa in water without MBs (Fig.~\ref{fig:IncidentFields}(a)). Compared to the x-wave case, the focused pressure field shows a more substantial level of attenuation below of the MB suspension with peak acoustic pressure at $(x,\ z)=(0,\ \mathrm{mm},\ 7\ \mathrm{mm})$ mm (b) of 140 kPa compared to 205 kPa in water without MBs (Fig.~\ref{fig:IncidentFields}(b)). Similarly, the planar pressure field shows a peak pressure of 186 kPa compared to 272 kPa in water without MBs (Fig.~\ref{fig:IncidentFields}(c)) and the diverging pressure field shows a peak pressure of 168 kPa compared to 243 kPa in water without MBs (Fig.~\ref{fig:IncidentFields}(d)).
		
		Residual AM pressure fields in the presence of a suspension of resonant monodisperse MBs are reported in Figs.~\ref{fig:TotalPressureFields}(e)-(h). For the xAM case, a better view of the spatial pattern of Fig.~\ref{fig:TotalPressureFields}(e) is illustrated in Fig.~\ref{fig:NonlinearInserts}(b) of Appendix ~\ref{app:ResidualFieldsInserts}. We observe that nonlinear effects accumulate along the direction of propagation of each plane wave. However, along the bisector and below the MB suspension, nonlinear effects are suppressed. The residual xAM pressure field reaches a maximum of $15.2\ \mathrm{kPA}$ within the MB suspension, as desired for a CEUS pulse sequence, and immediately drops beyond the MB suspension. A 1.7 kPa peak pressure is measured at $z=6.3$ mm.
		
		For the focused AM case (Fig.~\ref{fig:TotalPressureFields}(f)), the amplitude of the residual pressure field builds up inside the MB suspension along the $z$-axis and reaches a maximum of 51 kPa at $z$=5.7 mm, which is deeper than the transducer focus of 5.25 mm. Below the MB suspension, the amplitude of the residual AM pressure field remains high and reaches 47.7 kPa at $z=6.3$ mm, which is 28 times higher than for the xAM case.
		
		For the planar AM case (Fig.~\ref{fig:TotalPressureFields}(g)), the amplitude of the residual pressure field builds up along the $z$-axis and expands across the width of the plane wave beam profile. We observe an increase of nonlinear effects within and below of the MB suspension, which is as wide as the MB cloud. Cumulative nonlinear effects are also observed besides of the MB suspension, these are caused by nonlinear wave propagation alone. The planar AM pressure field reaches a peak of 52 kPa within the MB suspension and 53.4 kPa at $z=6.3$ mm, which is 31 times higher than for the xAM case.
		
		For the diverging AM case (Fig.~\ref{fig:TotalPressureFields}(h)), we observe an increase of nonlinear effects below the MB suspension that spans the full width of the MB suspension as well. The AM pressure field peaks at 58.6 kPa within the MB suspension and 59.4 kPa at $z=6.3$ mm, which is 35 times higher than for the xAM case.
		\begin{figure}[t!]
			\centering
			\includegraphics{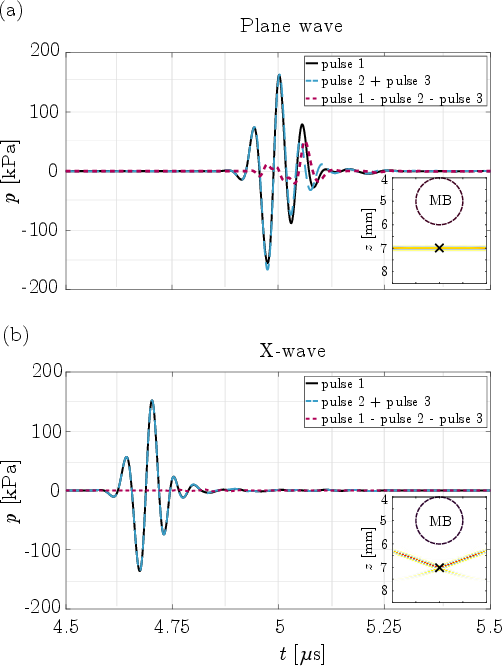}
			\caption{\label{fig:TS_PW_X} Pulses at a depth of 7 mm. (a) Waveforms for the planar AM pulse sequence. (b) Waveforms for the xAM pulse sequence.}
		\end{figure}
		%
		\subsection{Reason of reduction of nonlinear wave propagation artifacts by an x-shaped wavefront} \label{subsec:Results_Supersonic}
		%
		To elucidate why cumulative nonlinear effects do not build up along the bisector in the xAM case, in Fig.~\ref{fig:TS_PW_X} we present the transmitted waveforms with a planar wavefront and an x-shaped wavefront at a depth $z=7$ mm, which is below the MB suspension. 
		
		Fig.~\ref{fig:TS_PW_X}(a) displays the three pulses of the planar AM sequence at a depth of 7 mm. Pulses reach $z = 7$ mm in 4.73 $\mu$s (including 0.1$\mu$s delay), corresponding to a wave velocity of 1482 m$\cdot\mathrm{s}^{-1}$. The residual waveform derived from the AM operation exhibits a non-zero pressure amplitude with a residual AM peak pressure of 50.6 kPa compared to 162.2 kPa for the first pulse of the sequence. We observe that the sum of the second and third pulses of the sequence (Fig.~\ref{fig:TS_PW_X}(a), dashed magenta line) does not match the waveform of the double amplitude pulse 1 (Fig.~\ref{fig:TS_PW_X}a, continuous black line). 
		
		In comparison, Fig.~\ref{fig:TS_PW_X}(b) shows that the three pulses of the xAM sequence propagate to a depth of 7 mm in 4.45 $\mu$s (including 0.1$\mu$s delay), corresponding to a supersonic wave velocity of $c_\mathrm{X} =1577$ m$\cdot\mathrm{s}^{-1}$. The sum of the second and third pulses of the xAM sequence overlaps nearly perfectly with the first pulse. The residual waveform derived from the xAM operation exhibits a quasi-zero residual pressure amplitude of 1.7 kPa after the xAM operation compared to 152.1 kPa for the first pulse.
		\bigskip
		\bigskip
		%
		\subsection{Effect of the cross-propagation angle $\theta$ on the reduction of the nonlinear artifact} \label{subsec:Angle_dependency}
		%
		%
		\begin{figure}[bp!]
			\centering
			\includegraphics{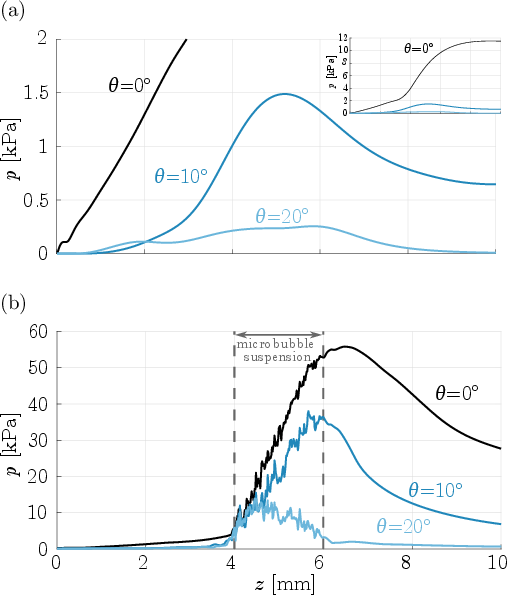}
			\caption{\label{fig:Angle_Dependency} xAM peak residual pressure along the bisector, as a function of depth for increasing angle $\theta$. (a) Residual pressure profiles in the absence of MBs.
				(b) Residual pressure profiles in the presence of monodisperse MBs. The boundaries of the MB suspension are indicated with dashed vertical lines.}
		\end{figure}
		\begin{figure*}[htbp!]
			\centering
			\includegraphics{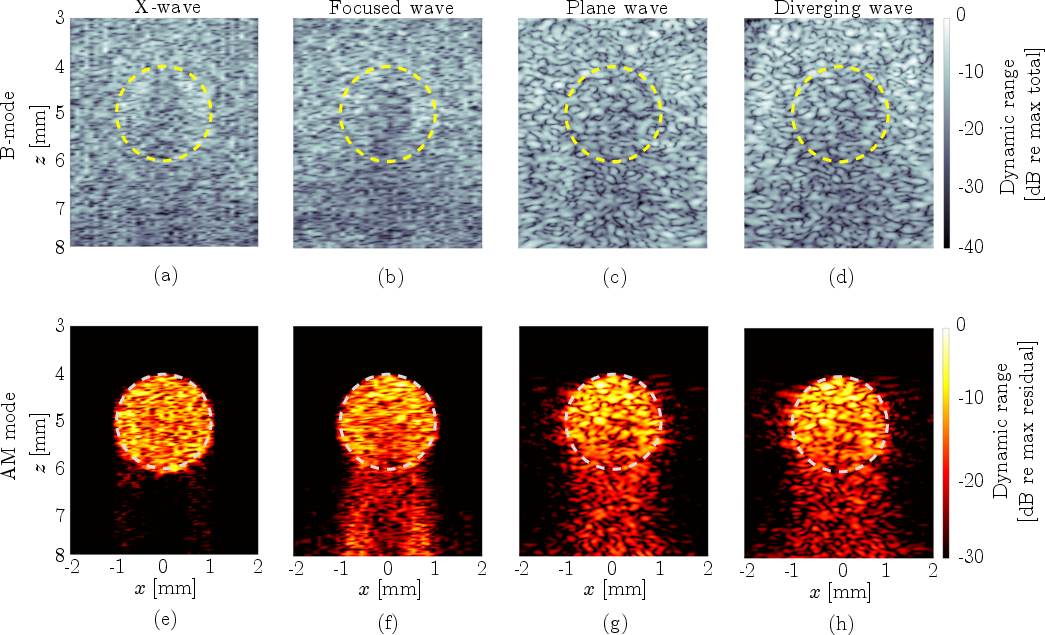}
			\caption{\label{fig:BeamformedImg} Numerical B-mode and AM ultrasound images of monodisperse MBs. (a)-(d) Single-shot ultrasound images acquired with x-shaped, focused, planar and diverging wavefronts, respectively. (e)-(h) AM ultrasound images acquired with x-shaped, focused, planar and diverging wavefronts, respectively. The position of the MB suspension is outlined by a dashed circle. The surrounding domain is filled with tissue-mimicking linear scatterers.}
		\end{figure*}
		Figs.~\ref{fig:Angle_Dependency}(a) and (b) display the peak of the AM residual pressure as a function of depth along the bisector, in the absence and presence of monodisperse MBs, respectively. Figure~\ref{fig:Angle_Dependency}(a) shows that higher angles $\theta$ are more effective at suppressing effects arising from nonlinear wave propagation in the medium.
		
		Figure~\ref{fig:Angle_Dependency}(b) shows that the presence of monodisperse MBs leads to a substantial increase in the residual xAM peak pressure level within the boundaries of the MB suspension, as desired for a CEUS imaging mode. However, we observe that angles of $\theta=0^{\circ}$ and $\theta=10^{\circ}$ do not strongly suppress nonlinear effects below the MB suspension. In contrast, an angle of$\theta=20^{\circ}$ strongly reduces the xAM pressure amplitude beyond the boundary of the MB suspension.
		%
		\subsection{Impact of wavefront shape on AM imaging of monodisperse MBs} \label{subsec:Results_Imaging}
		%
		In this section, we use INCS to simulate ultrasound imaging of a monodisperse MB suspension applying AM pulse sequences with four different wavefront shapes (Fig.~\ref{fig:ImagingSequences}). The computational domain is shown in Fig.~\ref{fig:ContrastDomain}(b). Conventional single-shot ultrasound images, also referred to as B-mode images, are reported in Figs.~\ref{fig:BeamformedImg}(a)-(d). The backscattered amplitude arising from tissue-mimicking linear scatters and resonant monodisperse MBs generates a similar echogenicity level, independent of the acoustic wavefront shape. As such, B-mode images do not allow to disentangle monodisperse MBs and tissue mimicking scatterers.
		
		AM images generated with each wavefront shape are reported in Figs.~\ref{fig:BeamformedImg}(e)-(h). The xAM sequence (Fig.~\ref{fig:BeamformedImg}(e)) generates a highly specific image of monodisperse MBs. On the contrary, AM sequences based on focused, planar and diverging wavefronts (Figs.~\ref{fig:BeamformedImg}(f)-(h)) generate images of MBs that suffer from significant nonlinear wave propagation artifacts below the MB suspension. The level of nonlinear artifacts is -30 dB for the xAM image, -9 dB for the focused AM image, -12 dB for the plane wave AM image and -12 dB for the diverging AM image.
		\bigskip
		\bigskip
		%
		\section{Discussion} \label{sec:Discussion}	
		%
		We report a numerical investigation of CEUS imaging and demonstrate that AM imaging based on x-shaped wavefronts enables specific and sensitive imaging of monodisperse MBs at a high ultrasound frequency. Our results reveal that the shape of transmitted ultrasonic wavefronts plays a critical role in the specificity of AM ultrasound imaging of monodisperse MBs. Widely used AM pulse sequence implementations based on focused, planar and diverging wavefronts (Figs.~\ref{fig:ImagingSequences}(b)-(d)) are all subject to significant nonlinear wave propagation artifacts appearing below monodisperse MB suspensions (Figs.~\ref{fig:BeamformedImg}(f)-(h)).
		
		The INCS method enables to disentangle different nonlinear effects involved in CEUS. In the absence of MBs, a first observation is that AM pulse sequences for all four wavefront shapes capture cumulative nonlinear effects caused by ultrasound wave propagation, despite the low acoustic pressure transmitted in the medium, which corresponds to a mechanical index of 0.074 in our study (Figs.~\ref{fig:IncidentFields}(f)-(h)).
		A second observation is that a resonant monodisperse MB suspension amplifies cumulative nonlinear wave propagation effects in the direction of wave propagation for all wavefront shapes investigated (Figs.~\ref{fig:TotalPressureFields}(e)-(h)). As a result, for the focused, planar and diverging wavefronts, the AM operation is compromised below the monodisperse MB suspension as the involved pulses do not cancel each other anymore (Fig.~\ref{fig:TS_PW_X}(a)). This amplitude-dependent nonlinear scattering effect leads to nonlinear wave propagation artifacts in AM images of monodisperse MBs (Figs.~\ref{fig:BeamformedImg}(f)-(h)).
		
		X-shaped wavefronts provide a solution to this problem. In the absence of MBs, residual xAM pressure fields prevent cumulative nonlinear wave propagation effects along the bisector, i.e. the imaging line, as seen in Fig.~\ref{fig:IncidentFields}(e). In the presence of monodisperse MBs, cumulative nonlinear effects are visible in the direction of propagation of each cross-propagating plane wave (see Fig.~\ref{fig:NonlinearInserts}(b) of Appendix ~\ref{app:ResidualFieldsInserts}), however nonlinear effects do not build up along the bisector. In xAM imaging, the segments of each wavefront that contribute to the double amplitude can only co-propagate for short periods of time. As a consequence, amplitude-dependent cumulative nonlinear effects are minimized and higher angles $\theta$ lead to a higher suppression of these effects (Fig.~\ref{fig:Angle_Dependency}).
		
		A limitation of our study is that we investigate AM ultrasound imaging of a perfectly monodisperse ultrasound contrast agent rather than an ultrasound contrast agent with a narrow size distribution. However, nonlinear artifacts observed in our study are therefore maximized and our study represents the worst-case test scenario for xAM imaging. Another limitation is that we only investigate perfectly symmetric geometries. In the future, it would be interesting to evaluate the effect of a scattering structure located in the path of one of the two cross-propagating plane waves. Nevertheless, in vivo experiments prove that xAM imaging is also performing well in non-symmetric soft biological tissue \cite{Hurt2023}.
		Another interesting pulse sequence to investigate would be pulse inversion. It is acknowledged that pulse inversion sequences are prone to artifacts \cite{Tang2006} since they were initially developed to isolate second harmonic generated by nonlinear wave propagation \cite{Averkiou2000, Shen2005}. 
		
		Our study also has implications for the field of ultrasound localization microscopy \cite{Heiles2022}. Approaches that rely on AM sequences with focused, planar or diverging wavefronts  \cite{Yan2023} are likely to be subject to artifacts, especially when mapping the coronary vasculature of the inferior myocardial wall which sits below the left ventricle chamber filled with MBs.
		%
		\section{Conclusions} \label{sec:Conclusions}
		%
		We numerically investigated AM ultrasound imaging of monodisperse MBs for implementations relying of x-shaped, focused, planar and diverging acoustic wavefronts. We show that xAM ultrasound imaging can detect monodisperse MBs, the most nonlinear ultrasound contrast agent to date, with high sensitivity and specificity. All other AM sequence implementations were prone to nonlinear wave propagation artifacts. This study paves the way for numerical investigations and optimization of CEUS imaging.
		%
		\begin{acknowledgments}
			This research was supported by the project "Optoacoustic sensor and ultrasonic microbubbles for dosimetry in proton therapy" of the Dutch National Research Agenda which is partly financed by the Dutch Research Council (NWO). R.~Waasdorp was supported by a grant from the Medical Delta Ultrafast Ultrasound Heart and Brain Program. We thank G.~Renaud and B.~Heiles for the fruitful discussions.
		\end{acknowledgments}
		%
		
		\appendix
		\appsection{Equalization of incident acoustic pressure levels}\label{app:PressureEq}
		%
		%
		\begin{figure}[bp!]
			\centering
			\includegraphics{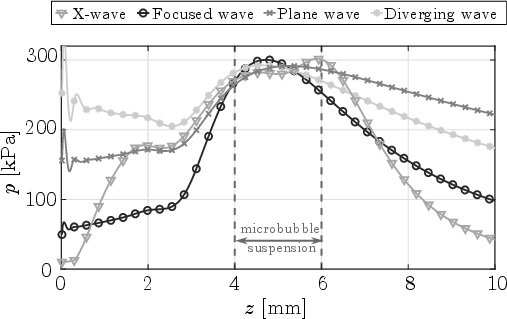}
			\caption{\label{fig:PressureEquation} Equalization of the average acoustic pressure transmitted by the first pulse of each sequence. Transmission for (a) x-wave (triangles), (b) focused wave (circles), (c) plane wave (xs) and (d) diverging wave (asterisks). The position of the MB suspension is indicated by the dashed vertical lines.}
		\end{figure}
		In this paper we compare AM sequences for four different wavefront shapes. These are obtained with varying numbers of active elements, different time delays and apodization, which may result in substantial differences in the generated incident pressure fields. To ensure a fair comparison of CEUS imaging results, it is necessary to deliver the same average peak pressure within the MB suspension for all cases. Furthermore, since image reconstruction for the x-wave and focused wave methods is performed in a line-by-line format, it is critical to equalize the average peak pressure along the bisector at the location of the MB suspension ($x=0,\ 4 \mathrm{mm}<z<6 \mathrm{mm}$). We have chosen to deliver an average peak acoustic pressure of 285.5 kPa over this interval for the first pulse of each AM pulse sequence. This corresponds to a mechanical index of 0.074. The resulting peak pressure as a function of depth along the bisector is given in (Fig.~\ref{fig:PressureEquation}) for each AM pulse sequence. 
		\begin{figure}[b!]
			\centering
			\includegraphics{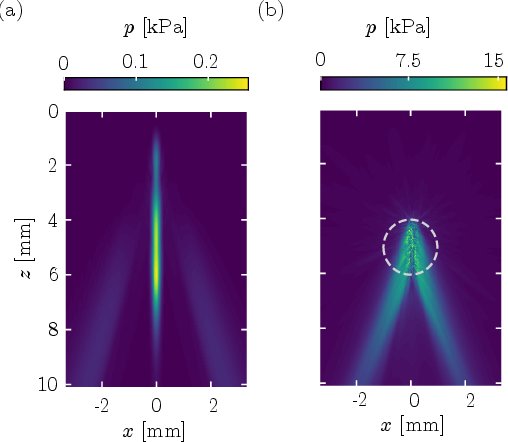}
			\caption{\label{fig:NonlinearInserts} Residual xAM pressure fields in (a) a homogeneous nonlinear medium consisting of water without MBs and (b) a medium consisting of water with nonlinear monodisperse MBs. Dashed white lines indicate the position of the MB suspension.}
		\end{figure}
		%
		\appsection{Residual xAM pressure fields in the absence and presence of monodisperse MBs}\label{app:ResidualFieldsInserts}
		To better visualize the residual xAM pressure fields reported in Figs.~\ref{fig:IncidentFields}(e) and \ref{fig:TotalPressureFields}(e), in Fig.~\ref{fig:NonlinearInserts} we show the same results using a smaller pressure range.
		
		

\begin{thebibliography}{10}
			\newcommand{\enquote}[1]{``#1''}
			\expandafter\ifx\csname url\endcsname\relax
			\def\url#1{\texttt{#1}}\fi
			\expandafter\ifx\csname urlprefix\endcsname\relax\def\urlprefix{URL }\fi
			\providecommand{\bibinfo}[2]{#2}
			\providecommand{\noopsort}[1]{}
			\providecommand{\switchargs}[2]{#2#1}
			
			\bibitem{Errico2015}
			\bibinfo{author}{C. Errico}, \bibinfo{author}{J. Pierre}, \bibinfo{author}{S. Pezet}, \bibinfo{author}{Y. Desailly}, \bibinfo{author}{Z. Lenkei}, \bibinfo{author}{O. Couture} and
			\bibinfo{author}{M. Tanter},
			\enquote{\bibinfo{title}\textit{{Ultrafast ultrasound localization microscopy for deep super-resolution vascular imaging}}},
			\href{https://doi.org/10.1038/nature16066}{\bibinfo{journal}{Nat. }\textbf{\bibinfo{volume}{527}}, \bibinfo{pages}{499--502}
				(\bibinfo{year}{2015}).}
			
			\bibitem{Heiles2021}
			\bibinfo{author}{B. Heiles}, \bibinfo{author}{D. Terwiel}, and
			\bibinfo{author}{D. Maresca},
			\enquote{\bibinfo{title}{\textit{The Advent of Biomolecular Ultrasound Imaging}}},
			\href{https://doi.org/10.1016/j.neuroscience.2021.03.011}{\bibinfo{journal}{Neurosc. }\textbf{\bibinfo{volume}{474}}, \bibinfo{pages}{122--133}
				(\bibinfo{year}{2021}).}
			
			\bibitem{Marmottant2005}
			\bibinfo{author}{P.~Marmottant}, \bibinfo{author}{S.~van~der~Meer}, \bibinfo{author}{M.~Emmer}, \bibinfo{author}{M.~Versluis}, \bibinfo{author}{N.~de~Jong},
			\bibinfo{author}{S.~Hilgenfeldt}, and \bibinfo{author}{D.~Lohse},
			\enquote{\bibinfo{title}{\textit{A model for large amplitude oscillations of coated bubbles accounting for buckling and rupture}}},
			\href{https://doi.org/10.1121/1.2109427}{\bibinfo{journal}{J. Acoust. Soc. Am.}
				\textbf{\bibinfo{volume}{118}}, 6, \bibinfo{pages}{3499--3506} (\bibinfo{year}{2005}).}
			
			\bibitem{Averkiou2020}
			\bibinfo{author}{M.~Averkiou}, \bibinfo{author}{M.~Bruce}, \bibinfo{author}{J.~Powers}, \bibinfo{author}{P.~Sheeran}, and \bibinfo{author}{P.~Burns},
			\enquote{\bibinfo{title}{\textit{Imaging methods for ultrasound contrast agents}}},
			\href{https://doi.org/10.1016/j.ultrasmedbio.2019.11.004}{\bibinfo{journal}{Ultrasound Med. Biol. \textbf{46}}, 3, \bibinfo{pages}{498--517} (\bibinfo{year}{2020}).}
			
			\bibitem{Tang2006}
			\bibinfo{author}{M.-X.~Tang}, and \bibinfo{author}{R.~Eckersley},
			\enquote{\bibinfo{title}{\textit{Nonlinear propagation of ultrasound through microbubble contrast agents and implications for imaging}}},
			\href{https://doi.org/10.1109/TUFFC.2006.189}{
				\bibinfo{journal}{IEEE Trans. Ultrason. Ferroelectr. Freq. Control  \textbf{53}, 12}, \bibinfo{pages}{2406--2415} (\bibinfo{year}{2006}).}
			
			\bibitem{tenKate2012}
			\bibinfo{author}{G.~ten~Kate}, \bibinfo{author}{G.~Renaud}, \bibinfo{author}{A.~Akkus}, \bibinfo{author}{S.~van den Oord}, \bibinfo{author}{F.~ten~Cate}, \bibinfo{author}{V.~Shamdasani}, \bibinfo{author}{R.~Entrekin}, \bibinfo{author}{E.~Sijbrands}, \bibinfo{author}{N.~de~Jong}, \bibinfo{author}{J.~Bosch}, \bibinfo{author}{A.~Schinkel}, and \bibinfo{author}{A.F.~van~der~Steen},
			\enquote{\bibinfo{title}{\textit{Far-wall pseudoenhancement during contrast-enhanced ultrasound of the carotid arteries: clinical description and in vitro reproduction}}},
			\href{https://doi.org/10.1016/j.ultrasmedbio.2011.12.019}{\bibinfo{journal}{Ultrasound Med. Biol. \textbf{38}}, 4, \bibinfo{pages}{593-600} (\bibinfo{year}{2012}).}
			
			\bibitem{Emmer2009}
			\bibinfo{author}{M.~Emmer}, \bibinfo{author}{H.J.~Vos}, \bibinfo{author}{D.E.~Goertz}, \bibinfo{author}{A.~van Wamel},  \bibinfo{author}{M.~Versluis}, and \bibinfo{author}{N.~de~Jong},
			\enquote{\bibinfo{title}{\textit{Pressure-Dependent Attenuationand Scattering of Phospholipid-Coated Microbubbles at Low Acoustic Pressures}}},
			\href{https://doi.org/10.1016/j.ultrasmedbio.2008.07.005}{\bibinfo{journal}{Ultrasound Med. Biol. \textbf{35}, 1}, \bibinfo{pages}{102--111} (\bibinfo{year}{2009}).}
			
			\bibitem{Sojahrood2023}
			\bibinfo{author}{A.~J.~Sojahrood}, \bibinfo{author}{Q.~Li}, \bibinfo{author}{H.~Haghi}, \bibinfo{author}{R.~Karshafian}, \bibinfo{author}{T.M.~Porter}, and
			\bibinfo{author}{M.C.Kolios}, \enquote{\bibinfo{title}
				{\textit{Probing the pressure dependence of sound speed and attenuation in bubbly media: Experimental observations, a theoretical model and numerical calculations}}}, 
			\href{https://doi.org/10.1016/j.ultsonch.2023.106319}{\bibinfo{journal}{Ultrason. Sonoch. \textbf{95}, 106319}, (\bibinfo{year}{2023}).}
			
			\bibitem{Renaud2015}
			\bibinfo{author}{G.~Renaud}, \bibinfo{author}{J.~Bosch}, \bibinfo{author}{A.F.~van~der~Steen},  and \bibinfo{author}{N.~de~Jong},
			\enquote{\bibinfo{title}{\textit{Increasing Specificity of Contrast-Enhanced Ultrasound Imaging Using the Interaction of Quasi Counter-Propagating Wavefronts: A Proof of Concept}}},
			\href{https://doi.org/10.1109/TUFFC.2015.007169}{
				\bibinfo{journal}{IEEE Trans. Ultrason. Ferroelectr. Freq. Control  \textbf{62}}, 10, \bibinfo{pages}{1768--1778} (\bibinfo{year}{2015}).}
			
			\bibitem{Maresca2018}
			\bibinfo{author}{D.~Maresca}, \bibinfo{author}{D.P.~Sawyer}, \bibinfo{author}{G.~Renaud}, \bibinfo{author}{A.~Lee-Gosselin}, and \bibinfo{author}{M.G.~Shapiro},
			\enquote{\bibinfo{title}{\textit{Nonlinear X-Wave Ultrasound Imaging of Acoustic Biomolecules}}},
			\href{https://doi.org/10.1103/PhysRevX.8.041002}{
				\bibinfo{journal}{Phys. Rev. X \textbf{8}, 041002} (\bibinfo{year}{2018}).}
			
			\bibitem{Nagai1982}
			\bibinfo{author}{S.~Nagai}, and \bibinfo{author}{K.~Iizuka},
			\enquote{\bibinfo{title}{\textit{A practical ultrasound axicon for non-destructive testing}}},
			\href{https://doi.org/10.1016/0041-624X(82)90047-6}{
				\bibinfo{journal}{Ultrasonics \textbf{20}, 6}, \bibinfo{pages}{265--270} (\bibinfo{year}{1982}).}
			
			\bibitem{Szabo2014}
			\bibinfo{author}{T.~L.~Szabo}, 
			\enquote{\bibinfo{title}{\textit{Diagnostic ultrasound imaging: Inside out}}},
			\href{https://doi.org/10.1016/C2011-0-07261-7}{
				\bibinfo{publisher}{Academic Press}
				(\bibinfo{year}{2014}).}
			
			\bibitem{Segers2018}
			\bibinfo{author}{T.~Segers}, \bibinfo{author}{P.~Kruizinga}, \bibinfo{author}{M.~Kok}, \bibinfo{author}{G.~Lajoinie}, \bibinfo{author}{N.~de~jong},  and \bibinfo{author}{M.~Versluis},
			\enquote{\bibinfo{title}{\textit{Monodisperse Versus Polydisperse Ultrasound Contrast Agents: Non-Linear Response, Sensitivity, and Deep Tissue Imaging Potential}}},
			\href{https://doi.org/10.1016/j.ultrasmedbio.2018.03.019}{\bibinfo{journal}{Ultrasound Med. Biol. \textbf{44}, 7}, \bibinfo{pages}{1482--1492} (\bibinfo{year}{2018}).}
			
			\bibitem{Tremblay2015}
			\bibinfo{author}{C.~Tremblay-Darveau}, \bibinfo{author}{R.~Williams}, \bibinfo{author}{L.~Milot}, \bibinfo{author}{M.~Bruce}, and \bibinfo{author}{P.~Burns},
			\enquote{\bibinfo{title}{\textit{Visualizing the Tumor Microvasculature With a Nonlinear Plane-Wave Doppler Imaging Scheme Based on Amplitude Modulation}}},
			\href{https://doi.org/10.1109/TMI.2015.2491302}{\bibinfo{journal}{IEEE Trans. Ultrason. Ferroelectr. Freq. Control \textbf{35}, 2}, \bibinfo{pages}{699--709} (\bibinfo{year}{2015}).}
			
			\bibitem{Yan2023}
			\bibinfo{author}{J.~Yan}, \bibinfo{author}{B.~Huang}, \bibinfo{author}{J.~Tonko}, \bibinfo{author}{M.~Toulemonde}, \bibinfo{author}{J.~Hansen-Shearer}, \bibinfo{author}{Q.~Tan}, \bibinfo{author}{K.~Riemer}, \bibinfo{author}{K.~Ntagiantas}, \bibinfo{author}{R.~Chowdhury}, \bibinfo{author}{P.~Lambiase}, \bibinfo{author}{R.~Senior},  and \bibinfo{author}{M.-X.~Tang}, \enquote{\bibinfo{title}{\textit{Transthoracic super-resolution ultrasound localisation microscopy of myocardial vasculature in patients}}},
			\href{https://doi.org/10.48550/arXiv.2303.14003}{
				\bibinfo{journal}{arXiv} (\bibinfo{year}{2023}).}
			
			\bibitem{Koos_Thesis}
			\bibinfo{author}{J.~Huijssen}, 
			\enquote{\bibinfo{title}{\textit{Modeling of nonlinear medical diagnostic ultrasound}}},\href{https://repository.tudelft.nl/islandora/object/uuid%3A3a01d973-d125-430f-82e2-fb83cc9239fb}{\bibinfo{journal} { Ph.D. Thesis, Delft University of Technology} (\bibinfo{year}{2008})}
			
			\bibitem{INCS}
			\bibinfo{author}{J.~Huijssen}, and \bibinfo{author}{M.D.~Verweij}, \enquote{\bibinfo{title}{\textit{An iterative method for the computation of nonlinear, wide-angle, pulsed acoustic fields of medical diagnostic transducers}}}, \href{https://doi.org/10.1121/1.3268599}{\bibinfo{journal}{J. Acoust. Soc. Am. \textbf{127}, 1}, \bibinfo{pages}{33--44} (\bibinfo{year}{2010}).}
			
			\bibitem{Libe_Thesis}
			\bibinfo{author}{L.~Demi}, 
			\enquote{\bibinfo{title}{\textit{Modeling nonlinear propagation of ultrasound through inhomogeneous biomedical media}}}, \href{https://repository.tudelft.nl/islandora/object/uuid:01b3942b-ffaa-4a27-be64-ea00f292bf5f/datastream/OBJ/download}{\bibinfo{journal}
				{Ph.D. Thesis, Delft University of Technology}
				(\bibinfo{year}{2013})}
			
			\bibitem{INCS_Inhom}
			\bibinfo{author}{L.~Demi}, \bibinfo{author}{K.W.A.~van~Dongen}, and \bibinfo{author}{M.D.~Verweij},
			\enquote{\bibinfo{title}{\textit{A contrast source method for nonlinear acoustic wave fields in media with spatially inhomogeneous attenuation}}}, 
			\href{https://doi.org/10.1121/1.3543986}{
				\bibinfo{journal}{J. Acoust. Soc. Am. \textbf{129}, 3}, \bibinfo{pages}{1221--1230} (\bibinfo{year}{2011}).}
			
			\bibitem{LocalNL2023}
			\bibinfo{author}{A.~Matalliotakis}, \bibinfo{author}{D.~Maresca}, and \bibinfo{author}{M.D.~Verweij}, \enquote{\bibinfo{title}{\textit{Nonlinear interaction of two cross-propagating plane waves}}}, \href{https://doi.org/10.48550/arXiv.2312.00445}{
				\bibinfo{journal}{arXiv} (\bibinfo{year}{2023}).}
			
			\bibitem{Bubble_Cloud}
			\bibinfo{author}{A.~Matalliotakis},  and
			\bibinfo{author}{M.D.~Verweij}, \enquote{\bibinfo{title}{\textit{Computation of ultrasound propagation in a population of nonlinearly oscillating microbubbles including multiple scattering}}}, 
			\href{https://doi.org/10.1121/10.0017770}{
				\bibinfo{journal}{J. Acoust. Soc. Am. \textbf{153}, 4}, \bibinfo{pages}{2209--2222} (\bibinfo{year}{2023}).}
			
			\bibitem{Segers2018a}
			\bibinfo{author}{T.~Segers}, \bibinfo{author}{E.~Gaud}, \bibinfo{author}{M.~Versluis}, and \bibinfo{author}{P.~Frinking},
			\enquote{\bibinfo{title}{\textit{High-precision acoustic measurements of the nonlinear dilatational elasticity of phospholipid coated monodisperse microbubbles}}},
			\href{https://doi.org/10.1039/C8SM00918J}{
				\bibinfo{journal}{Soft Matter}
				\textbf{\bibinfo{volume}{14}},
				\bibinfo{pages}{9550--9561}
				(\bibinfo{year}{2018}).}
			
			\bibitem{DAS2021}
			\bibinfo{author}{V.~Perrot}, \bibinfo{author}{M.~Polichetti},
			\bibinfo{author}{F.~Varray}, and
			\bibinfo{author}{D.Garcia}, \enquote{\bibinfo{title}
				{\textit{So you think you can DAS? A viewpoint on delay-and-sum beamforming}}}, \href{https://doi.org/10.1016/j.ultras.2020.106309}{\bibinfo{journal}{Ultrasonics \textbf{111 }, 106309}, (\bibinfo{year}{2021}).}
			
			\bibitem{MUSTToolbox}
			\bibinfo{author}{D.Garcia}, \enquote{\bibinfo{title}
				{\textit{Make the most of MUST, an open-source MATLAB UltraSound Toolbox}}}, \href{https://doi.org/10.1109/IUS52206.2021.9593605}{\bibinfo{journal}{IEEE Int. Ultrason. Symp.}, (\bibinfo{year}{2021}).}
			
			\bibitem{Lai2019}
			\bibinfo{author}{T.~Lai}, \bibinfo{author}{M.~Bruce}, and
			\bibinfo{author}{M.~Averkiou}, \enquote{\bibinfo{title}
				{\textit{Modeling of the Acoustic Field Produced by Diagnostic Ultrasound Arrays in Plane and Diverging Wave Modes}}}, 
			\href{https://doi.org/10.1109/TUFFC.2019.2908831}{\bibinfo{journal}{IEEE Trans. Ultrason. Ferroelectr. Freq. Control \textbf{66}}, 7, \bibinfo{pages}{1158--1169} (\bibinfo{year}{2019}).}
			
			\bibitem{Hurt2023}
			\bibinfo{author}{R.~Hurt}, \bibinfo{author}{M.~Buss}, \bibinfo{author}{M.~Duan}, \bibinfo{author}{K.~Wong}, \bibinfo{author}{M.~You}, \bibinfo{author}{D.~Sawyer}, \bibinfo{author}{M.~Swift}, \bibinfo{author}{P.~Dutka}, \bibinfo{author}{P.~Barturen-Larrea}, \bibinfo{author}{D.~Mittelstein}, \bibinfo{author}{Z.~Jin}, \bibinfo{author}{M.~Abedi}, \bibinfo{author}{A.~Farhadi}, \bibinfo{author}{R.~Deshpande}, and \bibinfo{author}{M.G.~Shapiro}, \enquote{\bibinfo{title}
				{\textit{Genomically mined acoustic reporter genes for real-time in vivo monitoring of tumors and tumor-homing bacteria}}}, 
			\href{https://doi.org/10.1038/s41587-022-01581-y}{
				\bibinfo{journal}{Nat. Biotech. \textbf{41}}, \bibinfo{pages}{919--931} (\bibinfo{year}{2023}).}
			
			\bibitem{Averkiou2000}
			\bibinfo{author}{M.~Averkiou}, \enquote{\bibinfo{title}{\textit{Tissue harmonic imaging}}}, \href{https://doi.org/10.1109/ULTSYM.2000.921622}{\bibinfo{journal}{IEEE Int. Ultrason. Symp.}, \bibinfo{pages}{1563--1572} (\bibinfo{year}{2000}).}
			
			\bibitem{Shen2005}
			\bibinfo{author}{C.-C. Shen}, 
			\bibinfo{author}{Y.-H. Chou}, 
			\bibinfo{author}{P.-C. Li}, 
			\enquote{\bibinfo{title}{\textit{Pulse inversion techniques in ultrasonic nonlinear imaging}}},
			\href{https://doi.org/10.1016/S0929-6441(09)60073-4}{\bibinfo{journal}{J. Med. Ultrasound} \textbf{\bibinfo{volume}{13}}, \bibinfo{number}{1}, \bibinfo{pages}{3--17}
				(\bibinfo{year}{2005}).}
			
			\bibitem{Heiles2022}
			\bibinfo{author}{B.~Heiles}, \bibinfo{author}{A.~Chavignon}, \bibinfo{author}{A.~Bergel}, \bibinfo{author}{V.~Hingot}, \bibinfo{author}{H.~Serroune}, \bibinfo{author}{D.~Maresca}, \bibinfo{author}{S.~Pezet}, \bibinfo{author}{M.~Pernot}, \bibinfo{author}{M.~Tanter}, and
			\bibinfo{author}{O.~Couture}, \enquote{\bibinfo{title}{\textit{Volumetric Ultrasound Localization Microscopy of the Whole Rat Brain Microvasculature}}}, 
			\href{https://doi.org/10.1109/OJUFFC.2022.3214185}{\bibinfo{journal}{IEEE Open Trans. Ultrason. Ferroelectr. Freq. Control  \textbf{2}}, \bibinfo{pages}{261--282} (\bibinfo{year}{2022}).} 
		\end{thebibliography}

	\end{document}